\documentclass{aastex62}

\received{}
\revised{}
\accepted{}
\submitjournal{ApJ}

\shorttitle{On the impact origin of Phobos and Deimos}
\shortauthors{Hyodo et al.}

\begin{document}

\title{On the Impact Origin of Phobos and Deimos I\hspace{-.1em}V: Volatile Depletion}

\correspondingauthor{Ryuki Hyodo}
\email{hyodo@elsi.jp}
\author{Ryuki Hyodo}
\affil{Earth-Life Science Institute/Tokyo Institute of Technology, 2-12-1 Tokyo, Japan}

\author{Hidenori Genda}
\affiliation{Earth-Life Science Institute/Tokyo Institute of Technology, 2-12-1 Tokyo, Japan}

\author{S{\'e}bastien Charnoz}
\affiliation{Institut de Physique du Globe/Université Paris Diderot 75005 Paris, France}

\author{Francesco C.F. Pignatale}
\affiliation{Institut de Physique du Globe/Université Paris Diderot 75005 Paris, France}

\author{Pascal Rosenblatt}
\affiliation{ACRI-ST, 260 route du pin-montard-BP 234, F-06904 Sophia-Antipolis Cedex, France}



\begin{abstract}
Recent works have shown that Martian moons Phobos and Deimos may have accreted within a giant impact-generated disk whose composition is about an equal mixture of Martian material and impactor material. Just after the giant impact, the Martian surface is heated up to $\sim3000-6000$ K and the building blocks of moons, including volatile-rich vapor, are heated up to $\sim2000$ K. In this paper, we investigate the volatile loss from the building blocks of Phobos and Deimos by hydrodynamic escape of vapor and radiation pressure on condensed particles. We show that a non-negligible amount of volatiles ($> 10\%$ of the vapor with temperature $> 1000$ K via hydrodynamic escape, and moderately volatile dusts that condense at $\sim700-2000$ K via radiation pressure) could be removed just after the impact during their first signle orbit from their pericenters to apocenters. Our results indicate that bulk Phobos and Deimos are depleted in volatile elements. Together with future explorations such as JAXA's MMX (Martian Moons eXploration) mission, our results would be used to constrain the origin of Phobos and Deimos.
\end{abstract}



\section{Introduction} \label{sec:intro}
The origin of Phobos and Deimos has been intensely debated. Historically, they were believed to be captured asteroids, due to their spectral properties sharing a resemblance with D-type asteroids \citep[e.g.][]{Bur78,Mur91}. However, the captured scenario is confronted with the difficulty of explaining their almost circular equatorial orbits around Mars \citep{Bur92,Ros11}. Recently, the giant impact scenario $-$ in which Phobos and Deimos accreted within an impact-generated disk $-$ is gaining more and more attention \citep{Ros16,Hes17,Hyo17a,Hyo17b,Hyo18}. Recent high-resolution smoothed-particles hydrodynamic (SPH) impact simulations show that the building blocks of Phobos and Deimos consist of a nearly equal mixture of Martian and impactor material \citep{Hyo17a}. They also found that a small amount of the building blocks is vaporized ($< 5$ wt\%) and the rest is melted ($> 95 $wt\%). Then, using the thermodynamic data obtained in \cite{Hyo17a}, \cite{Pig18} investigated the expected chemical composition of Phobos and Deimos assuming a variety of impactor compositions. They found that the vapor preferentially contains volatile elements and, during its condensation sequence, it morphs into different species depending on the impactor's composition. \cite{Ron16} investigated the formation of Mars' moons in an impact-generated disk that is composed of two main phases: a thin magma layer in the inner midplane and a larger gas envelope that extends at larger radii. After investigating the possible outcomes from magma solidification and gas condensation, \cite{Ron16} concluded that the condensed dust from the gas envelope in the outer region could be the origin of the Mars’ moons. However, they limit the study of condensability to olivine only, and thus, it is not possible to determine the full dust composition (amount of more or less volatiles species) that would be derived from their model.\\

In previous papers, the Martian moon-forming disk has been considered as a closed system. However, volatile elements in the vapor or condensed dust might preferentially escape from the system due to the fact that it is thermally energetic and the orbits of debris are highly eccentric just after the giant impact \citep{Hyo17a,Hyo17b}.\\
   
The volatile content of Martian moons would be an important proxy to reveal the origin of Phobos and Deimos, and JAXA (Japan Aerospace eXploration Agency) is planning the Martian Moons eXplorer (MMX) mission, in which a spacecraft will be sent to the Martian moons, conduct detailed remote sensing and in-situ analysis, retrieve samples there, and return to Earth.\\

In this paper, we consider hydrodynamic escape of volatile-rich vapor and removal of its condensates by planetary radiation pressure as possible mechanisms of volatile depletion from the building blocks of Phobos and Deimos. In section 2, we show that the surface of Mars is heated up significantly just after the impact and may be a major radiative source. In Section 3, we describe the spatial structure of the building blocks of Martian moons just after the impact. In section 4, we discuss the possibility of removing volatile-rich vapor by hydrodynamic escape. In section 5, we discuss the possibility of removing volatile-rich dust by planetary radiation pressure. In section 6 we summarize our results.

\section{Blazing Mars after a giant impact} \label{sec:surface}
\begin{figure}[ht!]
\plotone{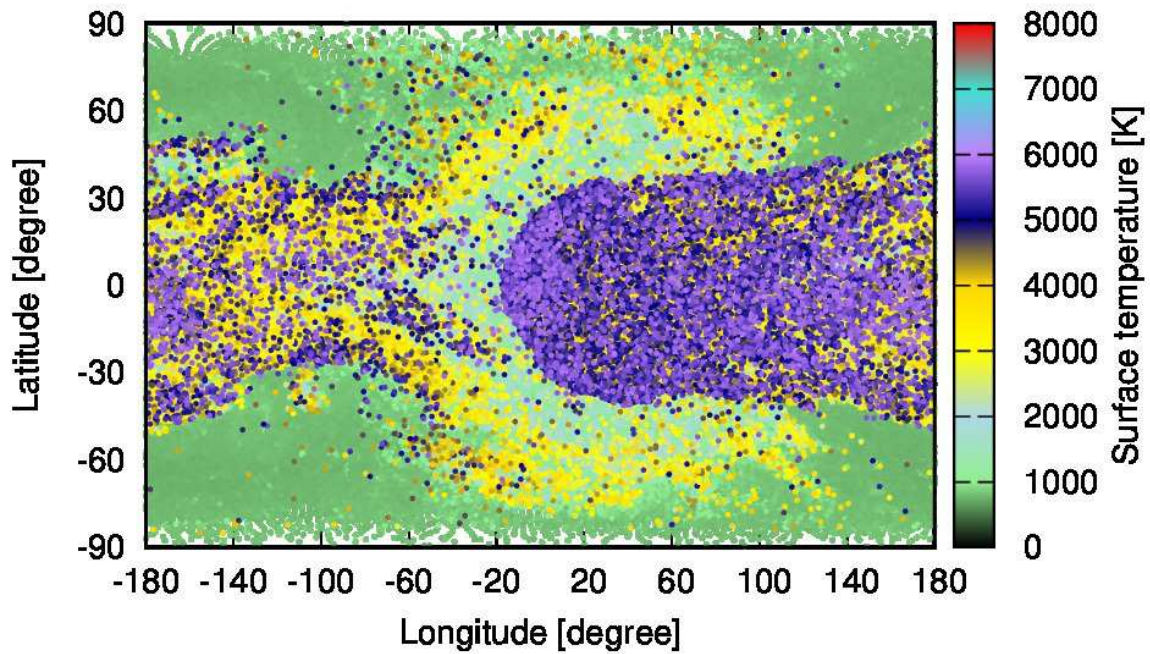}
\caption{Surface temperature distribution of Mars just after the Martian moon and Borealis basin forming impact obtained in \cite{Hyo17a}. Only particles whose depth is less than 100 km are over plotted (20h after the impact).}
\label{surface_temp}
\end{figure}

Using the data obtained from SPH impact simulations \citep{Hyo17a} that produce both the Martian moon-forming disk and the Borealis basin \citep{Mar08,Hyo17b}, we investigate the surface temperature of Mars just after the giant impact (impact energy of $\sim3-6 \times10^{29}$ J). Figure \ref{surface_temp} shows the post-impact temperature distribution of the Martian surface (impactor mass of $m_{\rm imp} \sim 0.03M_{\rm Mars}$, impact velocity of $v_{\rm imp} \sim1.4v_{\rm esc}$ where $v_{\rm esc}$ is the mutual escape velocity, and impact angle of $\theta=45$ degrees). We found that the impact significantly heats up the planet surface around the impact point. The post-impact Martian surface exhibits 3 distinct regions with temperatures $T_{\rm pla} \sim 5000-6000$ K, $\sim 3000-4000K$ and $\sim1000$ K. Note that the results do not significantly change when using other impact conditions that cover the similar impact energies ($m_{\rm imp}\sim0.01M_{\rm Mars}$, $v_{\rm imp}\sim2.2v_{\rm esc}$ and $\theta=45$ degrees, $m_{\rm imp}\sim0.056M_{\rm Mars}$, $v_{\rm imp}\sim1.4v_{\rm esc}$ and $\theta=45$ degrees).\\

A very simple estimation will help to quantify this temperature increase ($\Delta T$)
 \begin{equation}
 	\Delta T = E_{\rm heat}/C_{\rm p}M_{\rm heat}
 \end{equation}
where $E_{\rm heat}$ is the energy used to heat the Martian surface, $C_{\rm p}$ is the specific heat, and $M_{\rm heat}$ is the heated mass. Here, we consider $M_{\rm heat} \sim M_{\rm imp}$, because the volume of an isobaric core induced by the impact on Mars is comparable to that of the impactor. Since the total impact energy ($E_{\rm imp}$) is roughly equally partitioned to Mars and the impactor, and about half of this energy is used to increase the internal energy, we adopted $E_{\rm heat} \sim 0.25E_{\rm imp}$. We also adopted C$_{\rm p}$ = 1000 J K$^{-1}$ kg$^{-1}$ for typical rock. Then we get $\Delta T \sim 4000$ K, which is consistent with our numerical results.\\

The cooling timescale of the surface temperature anomaly can be estimated as follows. We consider the energy emitted by time unit 
\begin{equation}
	dE/dt = S \times \sigma_{\rm SB} T_{\rm pla}^4,
\end{equation}
and energy change 
\begin{equation}
	\Delta E = S \times D \times \rho C_{\rm p} \Delta T,
\end{equation}
where $\sigma_{\rm SB}=5.67 \times 10^{-5}$ (in cgs unit) is the Stefan-Boltzmann constant, $S$ is the surface area, $D$ is the depth and density $\rho=3000$ kg m$^{-3}$. Then, the cooling timescale can be written by assuming a black-body radiation cooling \citep[see also][]{Hyo17a} as 
 \begin{equation}
 		t_{\rm cool} = \Delta E/(dE/dt) \sim 717 \hspace{1mm} \rm{days} \times \left( \frac{D}{100 \rm{km}} \right) \left( \frac{\Delta T}{3000 \rm{K}} \right)  \left( \frac{T_{\rm pla}}{4000 \rm{K}} \right)^{-4}. 
 \end{equation}
Thus, it takes years to cool down from $T_{\rm pla}=4000$ K to $T_{\rm pla}=1000$ K assuming its depth of $100$ km (SPH simulations show that more than 100 km in depth is significantly heated up). Note that, in this work, we focus on the epoch just after the impact and before the disk particles are circularized to form a circular thin disk with a timescale of 10s of days \citep{Hyo17b}. Thus, the cooling timescale is much longer than the dynamical timescale.
  
\section{Disk structure just after the giant impact} \label{sec:disk_str}
  In this paper, we focus on the epoch just after the giant impact and before the debris were circularized to create the circular equatorial Martian moon-forming disk. This is because, just after the impact, the system is thermally hot and the debris particles have highly eccentric orbits \citep{Hyo17a,Hyo17b}. Thus, the particles can reach a location distant from Mars, where Martian gravitational attraction becomes weaker (closer to particles’ apocenter) and where escape of the planet is easier. Below, we describe the orbits and structure of the debris just after the impact where particles travel from their pericenter to apocenter distances.\\

\subsection{Orbits and configurations of disk particles}
\begin{figure}[ht!]
  \epsscale{0.50}
\plotone{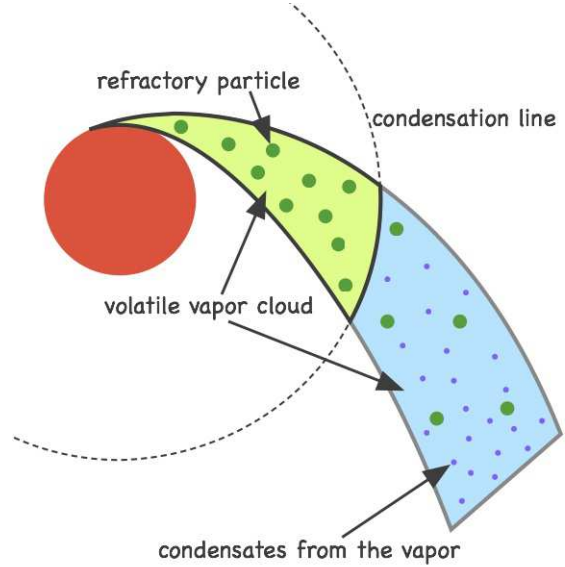}
\caption{Schematic figure of the building blocks of Phobos and Deimos just after the impact \citep[see also Figure 1 of][]{Hyo17a}. Closer to blazing Mars, the disk consists of vapor (light green region) and meter-sized melts (green points). In contrast, the outer part (outside the condensation line) consists of volatile-rich vapor (cyan region), meter-sized melts/solids (green points) and $\micron$-sized volatile-rich condensates from the vapor (small blue points).}
\label{disk_sch}
\end{figure}

\cite{Hyo17a} shows that the initial disk material just after the giant impact is mostly melted phase and $\sim 5$wt\% of the total disk mass ($M_{\rm tot} \sim 10^{20}$ kg) is vaporized. The orbits of the disk material are initially highly eccentric \citep[$e>0.5$,][]{Hyo17a,Hyo17b} and their radial distances significantly change with time. Thus, near the planet, the material may not be solid due to stronger radiative heating from blazing Mars, but as they approach their apocenter distance, they may solidify or condense (Figure \ref{disk_sch}). Typical particle size of the melts (or its solids) is $\sim 1.5$ m during their first orbit from their pericenters to apocenters \citep{Hyo17a}. Thus, such melt particles are too large to be blown off by Martian radiation pressure (see Section 5). In contrast, the condensates from the vapor are expected to have a typical size of $\sim0.1 \mu$m \citep{Ron16,Hyo17a} and thus they can potentially be removed by radiation pressure from blazing Mars.\\

\subsection{Temperature of particles heated by planetary radiation} \label{sec:disk_temp}
In this subsection, we estimate the temperature of particles as a function of radial distance from Mars. Here we consider the case where solid particles and ambient vapor quickly equilibrate. Below, we calculate the temperature of solids when they are put under planetary radiation. We assume that solids are heated up instantaneously and achieve an equilibrated temperature with radiation cooling.\\   

\begin{figure}[ht!]
\plotone{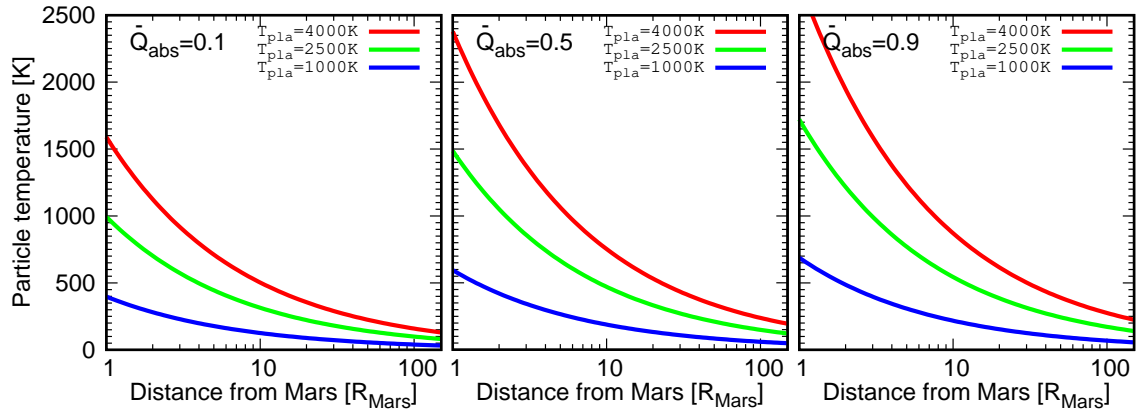}
\caption{Temperature of particles as a function of radial distance from Mars, assuming balance between radiation heating and radiation cooling. Different colors represent cases of Mars with different temperatures. From left to right panels, we assume different values of $\bar{Q}_{\rm abs}=0.1, 0.5, 0.9$, respectively.}
\label{disk_temp}
\end{figure}

The balance between the planetary radiation heating whose surface temperature of $T_{\rm pla}$ and radiation cooling of particles whose temperature and size are $T_{\rm par}$ and $d$, respectively, can be written as 
\begin{equation}
	\bar{Q}_{\rm abs} \frac{\sigma_{\rm SB} T_{\rm pla}^4 \times 4\pi R_{\rm p}^2}{4\pi r^2} \times \pi d^2 = \sigma_{\rm SB}T_{\rm par}^4 \times 4 \pi d^2 
\label{equi_temp}
\end{equation}
where $R_{\rm p}$ is the radius of the central planet, $r$ is the distance between the central planet and solid. $\bar{Q}_{\rm abs}$ is the absorption efficiency of a solid that determines the efficiency of absorbing the radiation as internal heating. Note that $\bar{Q}_{\rm abs}$ ranges from 0 to 1 and it strongly depends on the material properties and its size. Thus, in this paper, we treat this as a parameter and we use $\bar{Q}_{\rm abs} = 0.1,0.5$ and $0.9$ for reference. Using the equation \ref{equi_temp}, we can calculate the equilibrium temperature of the particles as
\begin{equation}
T_{\rm par} = \frac{T_{\rm pla}R_{\rm p}^{1/2} \bar{Q}_{\rm abs}^{1/4}}{(2 r)^{1/2}}.
\label{Tpar}
\end{equation}
Figure \ref{disk_temp} shows the temperature of solids (and equilibrated ambient vapor) as a function of distance from Mars, assuming different surface temperatures of Mars and $\bar{Q}_{\rm abs}$ obtained by using the equation \ref{Tpar}. Here, we assume that equilibrium quickly occurs. The temperature of optically thin dust is regulated by radiation cooling or planetary radiation heating. If a dust has a temperature above the equilibrium  temperature, the dust radiatively cools and its timescale of micron-size dust is very quick compared to the orbital timescale \citep{Hyo17a}. In contrast, if dust temperature is below the equilibrium temperature, the planetary radiation heats up the dust. Then, its timescale $t_{\rm heat}$ can be estimated using the same argument in Section 2 but with $dE/dt=\bar{Q}_{\rm abs} \frac{\sigma_{\rm SB} T_{\rm pla}^4 \times 4\pi R_{\rm p}^2}{4\pi r^2} \times \pi d^2$ ($\rho=3000$ kg m$^{-3}$ and $C_{\rm p}=1000$ J K$^{-1}$ kg$^{-1}$) as
 \begin{equation}
 		t_{\rm heat} = \Delta E/(dE/dt) \sim 3\times10^{-3} \hspace{1mm} \rm{s} \times \bar{Q}_{\rm abs}^{-1} \left( \frac{d}{0.1 \micron} \right) \left( \frac{\Delta T}{1000 \rm{K}} \right)  \left( \frac{T_{\rm pla}}{4000 \rm{K}} \right)^{-4} \left(\frac{r}{10 \rm{R_{\rm Mars}}} \right)^2. 
 \end{equation}
Therefore, the heating timescale is also much shorter than the orbital timescale and thus thermal equilibration would quickly occur during the first orbits of particles.\\

\cite{Hyo17a,Hyo17b} showed that particles have highly eccentric orbits just after the giant impact and their radial distance from Mars can change between $\sim 1-100 R_{\rm Mars}$ depending on their semi-major axis and eccentricity. Also, just after the impact (when all disk particles are significantly concentrated around the impact point), disk particles are optically thick and thus their temperature is regulated by impact-induced energy \cite[$T_{\rm par} \sim 2000$ K in][]{Hyo17a} and not by radiation heating discussed here. However, as the disk expands with cooling and if disk material goes far enough away from Mars, they become optically thin (see Section 5.1) and then these outer parts quickly reach their equilibrium temperature regulated by radiation heating. Therefore, the temperatures of gases and solids would change between $\sim 100 < T_{\rm par} < 2000$ K depending on the radial distance and $\bar{Q}_{\rm abs}$ (Figure \ref{disk_temp}).\\

\section{Volatile gas depletion by hydrodynamic escape} \label{sec:hydro}
   Just after the Martian moon-forming impact, the debris are a mixture of vapor ($\sim 5$wt\%) and magma ($\sim 95$wt\%) with a temperature of $\sim$2000 K \citep{Hyo17a} and vapor preferentially contains volatile elements \cite[Table 3 in][]{Pig18}. In this section, we estimate the amount of volatile gas that can be thermally removed from the building blocks of Phobos and Deimos.\\
   
\subsection{Volatile gas depletion from Martian moon-forming disk} 
If the vapor is thermally energetic enough compared to planet gravity, it would escape in a hydrodynamic manner, which is called the hydrodynamic escape. The ratio of the gravitational energy required for escape and the thermal energy of vapor is expressed using the escape parameter $\lambda_{\rm esc}$, \citep[e.g.][]{Gen03},
\begin{equation}
\lambda_{\rm esc} = GMm/kT_{\rm vap}r
\label{lambda_esc}
\end{equation}
where $G$ is the gravitational constant, $M$ is the mass of the central planet, $m$ is the mean molecular weight of vapor, $k$ is the Boltzmann constant and $T_{\rm vap}$ is the temperature of the vapor at a distance $r$ from the planet. If $\lambda_{\rm esc} < 1$ is satisfied, vapor can readily escape from the potential field of the planet since the thermal velocity is larger than the local escape velocity of the planet.\\

Here, we use the direct output obtained from SPH simulations provided by \cite{Hyo17a} (their Figure 6) and evaluate their $\lambda_{\rm esc}$ values during their first orbit from their impact point to their apocenter distances (that is just after the impact and we consider $r_{\rm peri} < r < r_{\rm apo}$ for each particle). This is because a chance to escape from the system would be the most likely during the first orbit (because they are eccentric and thermally energetic); the period before circularization and cooling down of the vapor throughout successive orbits \citep{Hyo17b}. Note that, \cite{Nak17} also considered hydrodynamic escape as a possible volatile depletion process of the Martian moon-forming disk. However, they consider the epoch after the debris were circularized to form a circular steady-state disk around Mars and, under this circumstance, $\lambda_{\rm esc} > 1$. Here, we consider the early epoch before the steady-state circular disk is formed when $\lambda_{\rm esc} < 1$.\\

Besides $T_{\rm vap}$ and $r$, $\lambda_{\rm esc}$ depends on the mean molecular weight ($m$), which is calculated as follows. The vapor contains a mixture of about one half Martian material and about one half impactor material \citep{Hyo17a}. \cite{Pig18} calculated the vapor composition at $2000$ K, assuming the impactor’s composition is of either Mars, CV, CI, EH or comet-like material. Here, using the same procedure as \cite{Pig18}, we calculated the mean molecular weight of the vapor at 2000K, 1500K, and 1000K during their condensation sequence using various impactor compositions.\\

\begin{figure}[ht!]
\plotone{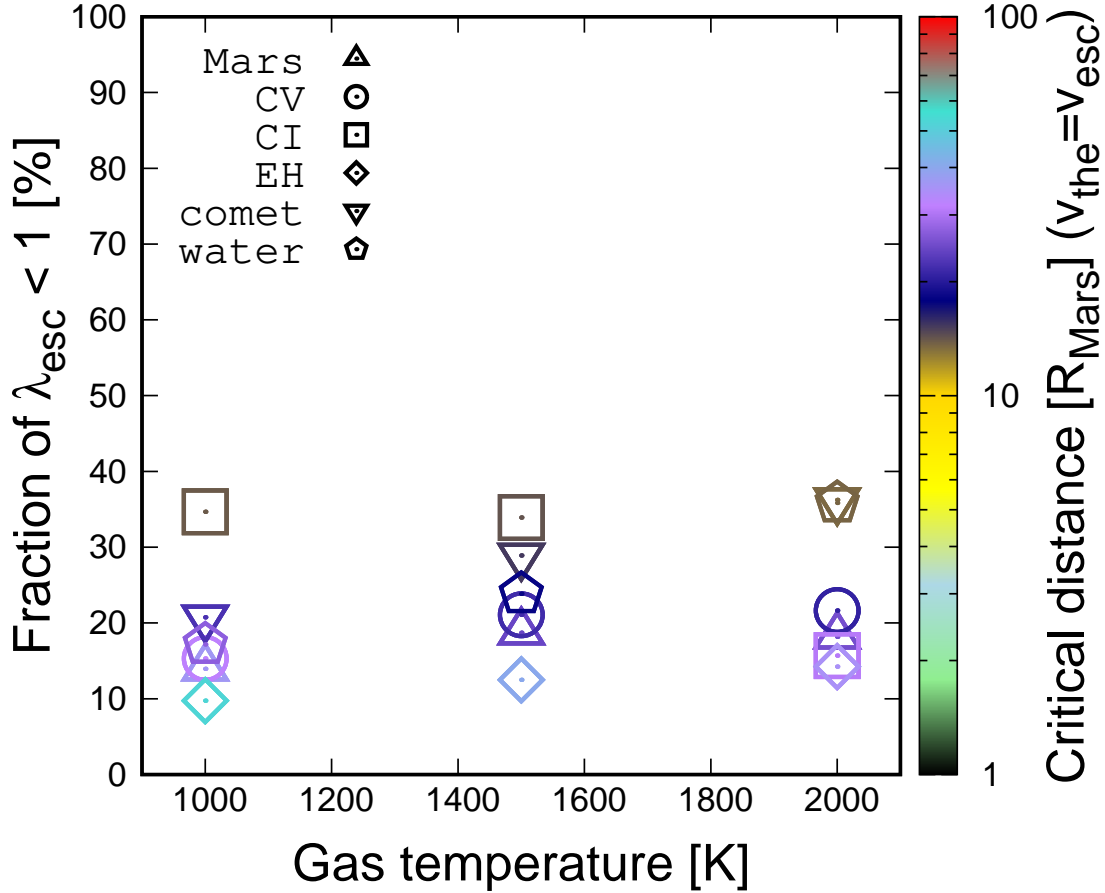}
\caption{Mass fraction of vapor phase that satisfies $\lambda_{\rm esc} < 1$ during the orbit from pericenter to apocenter using the data obtained in SPH simulations \citep{Hyo17a}. The mean molecular masses at different temperatures are obtained by calculating the condensation sequence starting from T=2000K and $P=10^{-4}$ bar, whose initial composition is the result of an equal mixture of Martian material and different impactor materials (Mars, CV, CI, EH or comet-like materials). As temperature decreases, the mean molecular mass becomes smaller since only more volatile elements remain in the vapor phase. The color contour represents the critical distance where $\lambda_{\rm esc} = 1$.}
\label{frac_hydro}
\end{figure}

Figure \ref{frac_hydro} shows the fraction of $\lambda_{\rm esc} < 1$ among eccentric particles during their first orbit as explained above. We found that a significant amount of about $\sim10-40$\% of the vapor (depending on the impactor composition) can hydrodynamic escape from the system when vapor temperature is 2000 K. Even when vapor temperature decreases down to 1000 K, $\sim10-40$\% of the vapor can still escape because the mean molecular weight decreases as temperature decreases (only more volatile elements remain in vapor phase).\\

If there is sufficient water on Mars at the time of impact, the impact-induced vapor may consist of mostly water ($m\sim18$ g mol$^{-1}$). In this case, about $\sim20-40\%$ of water-dominated vapor will be lost at a vapor temperature between $1000-2000$ K (see pentagon in Figure \ref{frac_hydro}).\\

Note that our above estimation only considers thermal velocity (using $\lambda_{\rm esc} < 1$ criteria) so that molecules are on ballistic trajectories. In reality, in addition to this thermal velocity, vapor itself has dynamical velocity inherited from the impact velocity (close to local Keplerian velocity) and a pressure gradient that reduces the effective gravity of Mars. These effects would ease the criteria of vapor escape from the Martian system and thus our above estimation should be considered as a lower bound.\\

\subsection{Mass fractionation through hydrodynamic escape} 
Mass fractionation, such as the change in $D/H$, is not expected for the rapid hydrodynamic escape considered here. The degree of mass fractionation can be evaluated from the crossover mass ($m_{\rm c})$, that is the mass of the heaviest species that can be dragged to space by the escaping species \citep{Hun87,Gen08} ;
\begin{equation}
m_{\rm c} = m + \frac{F_{esc}k T_{\rm vap}}{gb}
\label{}
\end{equation}
where $g$ is the gravitational acceleration, $F_{\rm esc}$ is the escape flux of major species, and $b$ is the binary diffusion coefficient. If $m_{\rm c}$ is comparable to $m$, mass fractionation is significant, while if $m_{\rm c} \gg m$ it is negligible. Here for simplicity, we consider a water-dominated vapor in the disk, i.e., $m = 3.0 \times10^{26}$ kg (equivalent to 18 g mol$^{-1}$). The escape flux Fesc is roughly estimated as
\begin{equation}
F_{\rm esc} \sim \frac{M_{\rm esc}}{mS \Delta t}
\label{}
\end{equation}
where $M_{\rm esc}$ is the total escaping mass ($\sim 10^{18}$ kg, which corresponds to 30\% of the vaporized disk mass), $S$ is the escaping surface area ($\sim 10^{14}$ m$^{2}$, which corresponds to the surface area of Mars), and $\Delta t$ is the typical duration of hydrodynamic escape. When we consider one orbit of the disk ($\Delta t \sim 1$ day), $F_{\rm esc}$ is estimated to be $10^{24}$ m$^{-2}$ s$^{-1}$. Since the order of magnitude for $b$ is $10^{22}$ m$^{-1}$ s$^{-1}$ for $T_{\rm vap} = 2000$ K \citep{Mas70}, and $g \sim 1$ m s$^{-2}$,
\begin{equation}
m_{\rm c}/m \sim 10^8
\label{}
\end{equation}
Therefore, no mass fractionation would take place during hydrodynamic escape in the Martian moon-forming disk.\\

\section{Volatile dust depletion by radiation pressure} \label{sec:RP}
In the previous section, we consider vapor loss by hydrodynamic escape. However, as shown in Section 2, the temperature of the vapor may decrease as the radial distance increases and previous papers show that the vapor may condense into $0.1 \mu m$ sized dust particles \citep{Hyo17a}. These small specks of dust may be affected by planetary radiation pressure. In this section, we investigate the possibility for such dust to be blown off by planetary radiation pressure just after the impact.\\

\subsection{Opacity of the debris disk} 
\begin{figure}[ht!]
\plotone{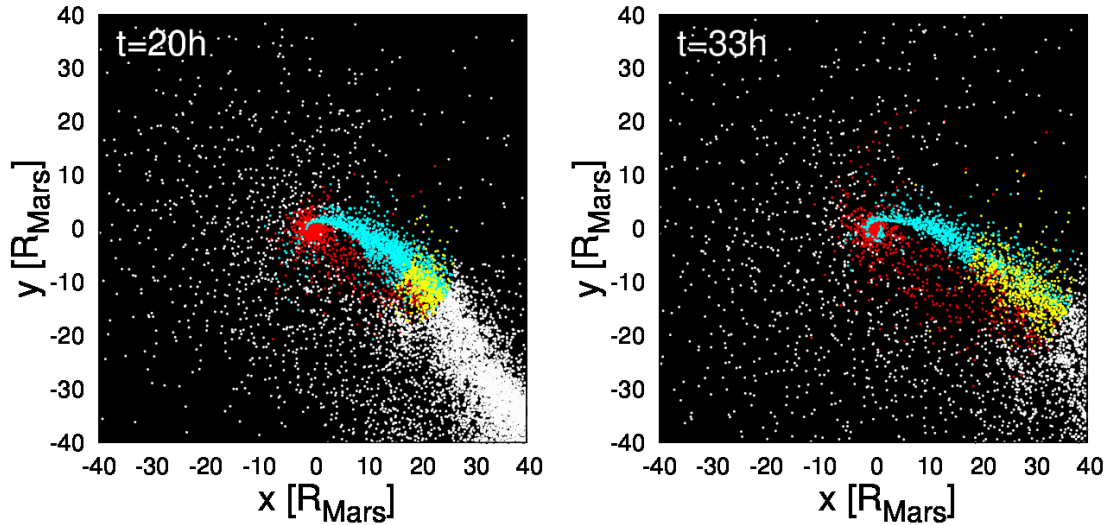}
\caption{Snapshots of impact simulations obtained in \cite{Hyo17a} (left panel is that of 20h and right panel is that of 33h). The red points represent those belonging to Mars. White points represent those that escaped from Mars’ gravity. Cyan points represent those of disk particles whose $\tau > 1$. Yellow points represent those of disk particles whose $\tau < 1$. About $\sim 20\%$ and $\sim 34$\% of the disk particles are $\tau < 1$ (yellow points) at 20h and 33h, respectively.}
\label{tau}
\end{figure}

Radiative escape of dusty volatile material would only be possible if the planet's hot and radiative surface is visible by the volatile condensates. To check for this possibility, we have computed the radially integrated optical depth ($\tau$) of the disk for every disk particle, using outputs of the SPH simulation at 20 h and 33h as examples. The system was divided into $1000\times200\times600$ cells in spherical coordinates with minimum radius at the planet surface and maximum radius at 500,000 km. Then, the mass in each cell was obtained by summing the mass of all particles inside a given cell and converting into an equivalent optical depth, assuming that all the mass is made of particles with a 1.5m radius \citep{Hyo17a}. Then the optical depth was computed for every cell along the radial path starting from Mars' surface. Results are shown in Figure \ref{tau}. It appears that particles close to Mars and accumulated in a dense tidal arm all have $\tau > 1$ and then should not be sensitive to radiation pressure. Conversely, all disk particles above and below this arm and beyond 15 Mars radii have $\tau < 1$, so will be subject to radiation pressure. These particles comprise about $\sim 20$\% (at 20h) and $\sim34\%$ (at 30h) of the total disk mass. Of course, as the system evolves, this fraction of particles with $\tau < 1$ increases as the systems spread radially and azimuthally. In addition, as these particles are located far from Mars, they are more prone to condense into small condensates. So these results suggest that a substantial population of small condensates and volatile particles will indeed be subject to radiative effects. Quantifying this number more precisely is a very difficult task as it would require a dynamical simulation fully coupled with a radiative transfer code, which is beyond the scope of the present paper that investigates first order effects.\\

\subsection{Basics of the radiation pressure} \label{sec:RP_basic}
The orbits of small particles may be significantly influenced by the radiation pressure either from the Sun or the central planet \citep[e.g.][]{Bur79}. The radiation pressure can be written as 
\begin{equation}
	F_{\rm RP} = \bar{Q}_{\rm RP} \frac{S}{c} \times \sigma_{\rm col}  		
\label{}
\end{equation}
where $\bar{Q}_{\rm RP}$ is the Planck mean of the radiation pressure efficiency averaged over spectrum, $S$ is the radiation flux density at distance $r$, $c$ is the speed of light, and $\sigma_{\rm col}=\pi d^2$ is the cross-section of a particle whose radius is $d$, respectively.\\

$\bar{Q}_{\rm RP}$ can be expressed using the radiation pressure efficiency $Q_{\rm RP}(\lambda, d)$ as a function of wavelength $\lambda$ and dust size $d$ as \citep{Bur79}
\begin{equation}
	\bar{Q}_{\rm RP}(T_{\rm pla},d)= \int_0^{\infty} B(\lambda,T_{\rm pla}) Q_{\rm RP} (\lambda,d) d\lambda		
\label{}
\end{equation}
where $B(\lambda, T_{\rm pla})$ is the normalized Planck function at a wavelength of $\lambda$ and planet temperature $T_{\rm pla}$ so that the total area of this curve is unity. Using the luminosity $L=\sigma_{\rm SB}T_{\rm pla}^4 \times 4\pi R_{\rm p}^2$, $S$ is written as $S=L/(4\pi r^2)$ and thus
\begin{equation}
	S=\frac{\sigma_{\rm SB} T_{\rm pla}^{4} \times 4\pi R_{\rm p}^2}{4\pi r^2}.	
\label{}
\end{equation}

As shown in section 2, the planet surface is significantly heated up by the giant impact. Assuming both the Sun and a heated planet emit black body radiation and assuming $\bar{Q}_{\rm RP}=1$, the ratio of these radiation pressures can be written as
\begin{equation}
	\frac{F_{\rm RP,Sun}}{F_{\rm RP,planet}} = \left( \frac{R_{\rm Sun}}{R_{\rm pla}} \right)^2 \left( \frac{r_{\rm pla}}{r_{\rm Sun}} \right)^2 \left( \frac{T_{\rm Sun}}{T_{\rm pla}} \right)^4   
\label{}
\end{equation}
where $R_{\rm Sun}=6.95 \times 10^5$ km and $R_{\rm pla}$ are the radii of the Sun and the planet, respectively. $r_{\rm Sun}=2.27 \times 10^8$ km and $r_{\rm pla}$ are the distances from the disk particles to the Sun and the planet, respectively. $T_{\rm Sun}$ and $T_{\rm pla}$ are the temperatures of the Sun and the planet, respectively. For the parameters of interest here, assuming $T_{\rm Sum}=6000$ K, $T_{\rm Mars}=3000K$ and $R_{\rm Mars}=3300$ km and $r_{\rm Mars}=4R_{\rm Mars}$, we get $F_{\rm RP,Sun}/F_{\rm RP,Mars} \sim 2\times10^{-3}$, which indicates the radiation pressure from Mars’ surface dominates over that from the Sun. Thus, in this paper, we only consider the effect of planetary radiation pressure on the building blocks of Phobos and Deimos.\\

\subsection{Conditions for eccentric dust to be blown off by radiation pressure} \label{sec:RP_cond}
\begin{figure}[ht!]
\plotone{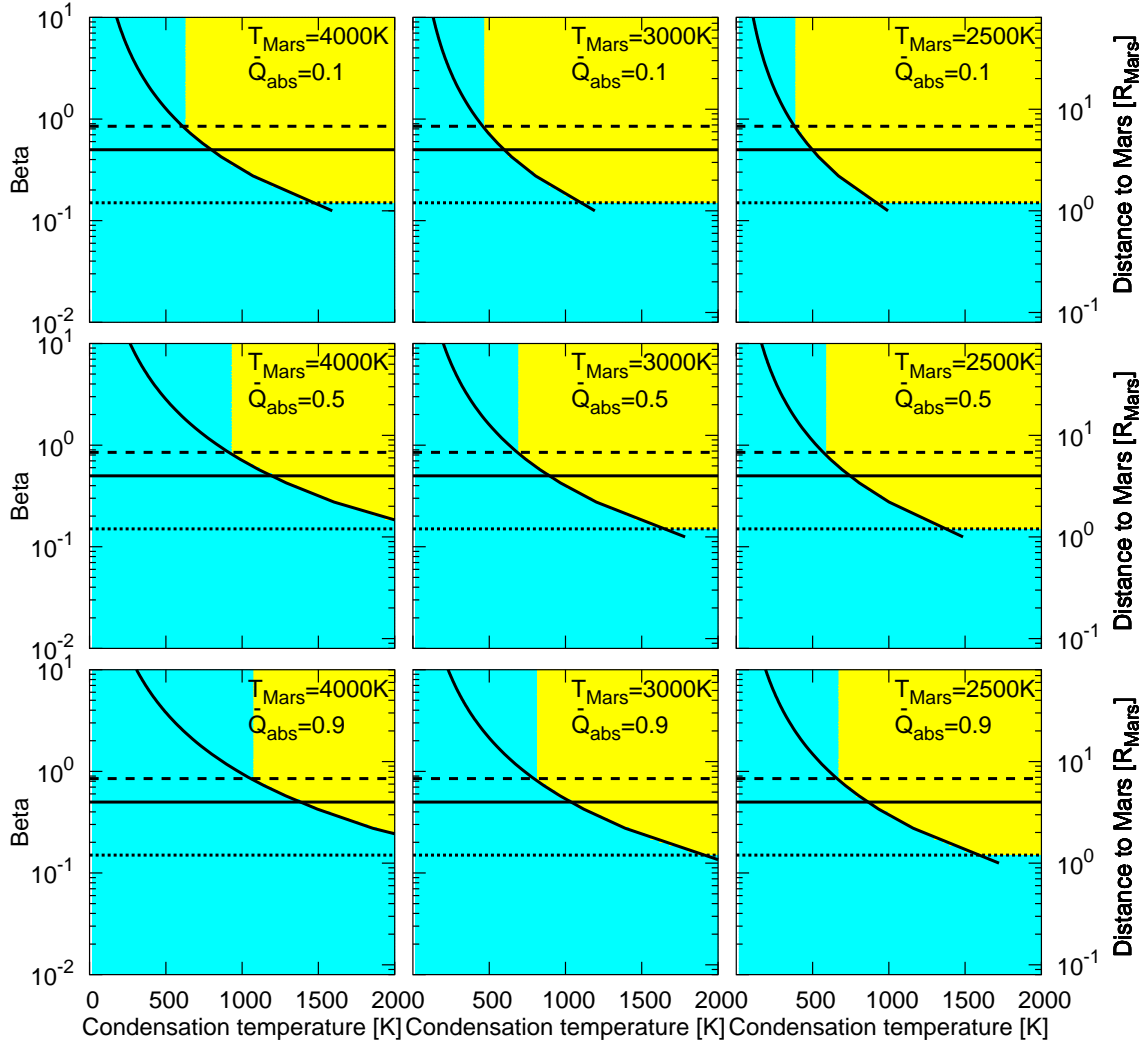}
\caption{An example of a map of parameters where a particle is removed by radiation pressure (yellow region). Cyan region is where the particle is not removed. The solid black curve is where particles are condensed, assuming particle temperature is settled by radiation equilibrium. The solid, dashed and dotted horizontal lines are the particle’s semi-major axis (corresponding to $\beta=0.5$), apocenter distance and pericenter distance, respectively. Note that, $a=4R_{\rm Mars}$ and $e=0.7$ are typical orbital elements found in SPH simulations \citep{Hyo17a,Hyo17b}. In Figure \ref{frac_RP}, we consider all the particles that have different orbital elements.}
\label{map}
\end{figure}

The vector of the radiation pressure points in the opposite direction of the gravitational force and thus the effective mass of the central planet can be expressed by using the ratio between these two absolute terms $\beta$ as
\begin{equation}
	M_{\rm eff} = \left(1-\beta \right)M 
\label{}
\end{equation}
where $\beta$ is written as
\begin{equation}
	\beta = \frac{F_{\rm RP}}{F_{\rm grav}} 
\label{}
\end{equation}
where $F_{\rm grav}=\frac{GM}{r^2}$.\\

In order for a condensed “eccentric” dust particle to be blown off by radiation pressure, the particle needs to have a larger velocity $v(r)$ than the escape velocity of the planet $v_{\rm esc}(r)$ at its distance r. We can write this condition as
\begin{equation}
	v(r)=\sqrt{GM\left( \frac{2}{r} - \frac{1}{a_{0}} \right) } > v_{\rm esc} = \sqrt{ \frac{2GM_{\rm eff}}{r}}
\label{}
\end{equation}
where $a_{\rm 0}$ is the semi-major axis of the dust particle. Since $r > r_{\rm peri}$, the critical blow-off conditions can be written as
\begin{equation}
	r_{\rm peri} < r < r_{\rm \beta} = 2 a_{0} \beta.
\label{}
\end{equation}
This criteria tells us that a particle on a circular orbit ($r=a_{\rm 0}$) requires $\beta=0.5$ as a critical value to be blown off \citep{Bur79}. However, if the orbit is eccentric, the critical $\beta$ value above which a particle is blown off depends on the radial location where particles condense in a way that a distance larger than $a_{\rm 0}$ requires $\beta > 0.5$ and a distance smaller than $a_{\rm 0}$ requires $\beta<0.5$.\\

In addition to the above criteria, a particle needs to condense (at $r=r_{\rm con}$) before they reach the apocenter to feel radiation pressure. Assuming the temperature of a particle is settled as an equilibrium temperature (Section \ref{sec:disk_temp}), the condition is written as 
\begin{equation}
	r_{\rm apo} > r > r_{\rm con} = \frac{1}{2} \left( \frac{T_{\rm pla}}{T_{\rm con}} \right)^2 R_{\rm p}  \bar{Q}^{1/2}_{\rm abs}. 
\label{}
\end{equation}
where $T_{\rm con}$ is the condensation temperature for a specific element. To satisfy the above two conditions at the same time, $r_{\rm \beta}$ and $r_{\rm con}$ need to be
\begin{equation}
	{\rm (i)\hspace{0.1cm}} r_{\rm con} < r_{\rm apo},  {\rm \hspace{0.2cm} (ii) \hspace{0.1cm}}  r_{\rm \beta} > r_{\rm peri},  \hspace{0.2cm} {\rm and \hspace{0.1cm} (iii) \hspace{0.1cm}}  r_{\rm con} < r_{\rm \beta}.
\label{}
\end{equation}
\\

Figure \ref{map} shows an example of a parameter map where a dust particle ($a=4R_{\rm Mars}$ and e=0.7) is removed by radiation pressure (yellow region) as a function of condensation temperature and $\beta$ at different planet temperatures and absorption coefficients of the particle. The sharp left vertical edge of the yellow region is due to the condition (i). Thus, particles whose condensation temperature is smaller than the critical temperature
\begin{equation}
	T_{\rm con,apo} = \sqrt{\frac{R_{\rm p}}{2r_{\rm apo}} } T_{\rm pla}  \bar{Q}^{1/4}_{\rm abs}
\label{Tapo}
\end{equation}
are never removed by radiation pressure. The horizontal bottom edge of the yellow region is due to condition (ii). The left bottom curves of the yellow regions represent condition (iii), which is the equilibrium temperatures of particles (minimum condensation temperatures) at a distance $r$ from Mars.\\

\subsection{$\beta$ value of the condensed dust particles} 
\begin{figure}[ht!]
\plotone{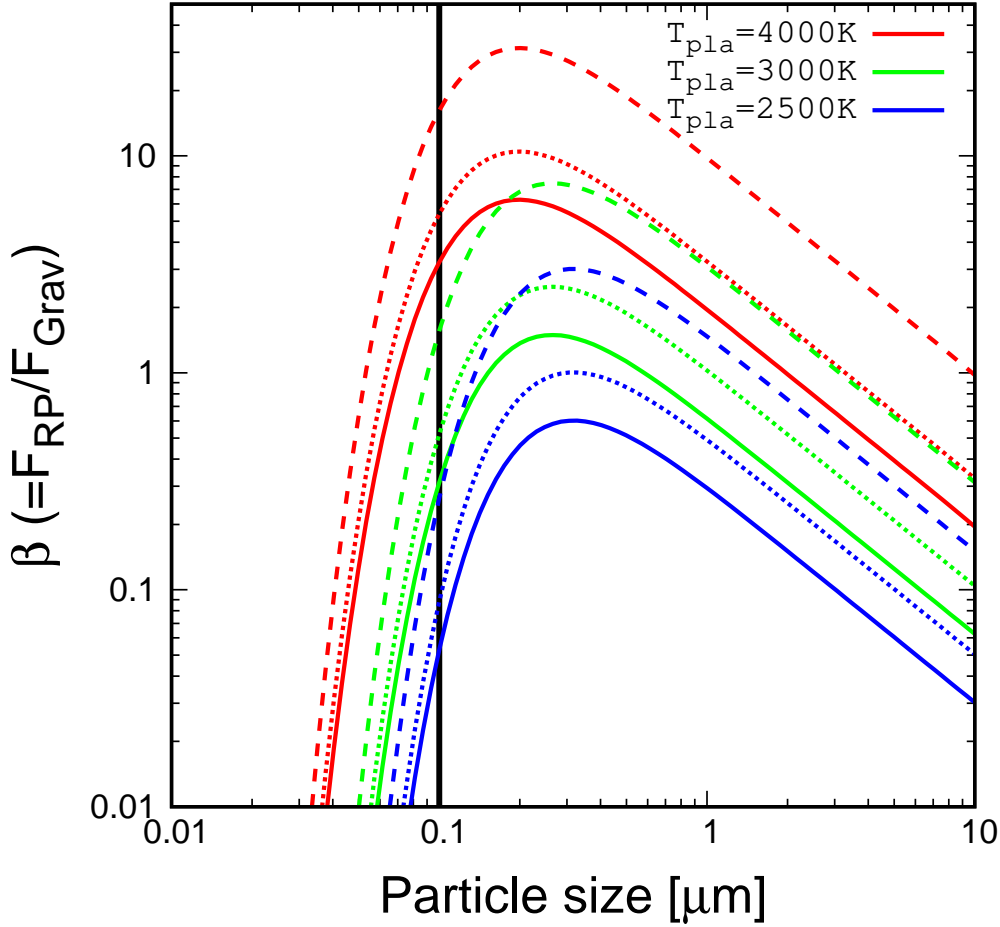}
\caption{$\beta$ value as a function of particle size at different planet temperatures and particle densities. Red, green and blue colors represent cases where planet temperatures are $T_{\rm pla}=4000$, 3000 and 2500 K, respectively. The solid, dotted and dashed lines represent cases where particle densities are $\rho_{\rm dust}=$5.0, 3.0 and 1.0 g cm$^{-3}$, respectively.}
\label{beta}
\end{figure}

In this subsection, we estimate the $\beta$ value of an arbitrary condensed species from the vapor. Following the previous discussion in Section \ref{sec:RP_basic}, $\beta$ can be written as
\begin{eqnarray}
\label{betavalue}
	&\beta = F_{\rm RP}/F_{\rm grav} =  \frac{9}{16\pi} \bar{Q}_{\rm RP} \times \frac{ \sigma_{\rm SB} T_{\rm pla}^4}{c G R_{\rm p} d \rho_{\rm pla} \rho_{\rm dust} }\\ 
	&\sim  10 \hspace{0.1cm}  {\rm [cgs]} \times  \bar{Q}_{\rm RP} \left( \frac{R_{\rm p}}{R_{\rm Mars}} \right)^{-1} \left( \frac{d}{0.1 \hspace{0.1cm}  \micron} \right)^{-1} \left( \frac{\rho_{\rm pla}}{ {\rm 4.0 \hspace{0.1cm} g \hspace{0.1cm} cm^{-3}} }\right )^{-1}  \left(\frac{\rho_{\rm dust}}{ {\rm 3.0 \hspace{0.1cm} g \hspace{0.1cm} cm^{-3} }} \right)^{-1}  \left( \frac{T_{\rm pla}}{3000 \hspace{0.1cm}  {\rm K}} \right)^4  
\end{eqnarray}
where $\rho_{\rm pla}$ and $\rho_{\rm dust}$ are densities of the central planet and irradiated particle, respectively. And now we need to evaluate the efficiency of radiation pressure $\bar{Q}_{\rm RP}$. Here, using the procedure discussed in \cite{Zoo75}, we simply estimate $\bar{Q}_{\rm RP}$ and calculate the $\beta$ value. We assume that $Q_{\rm RP}(\lambda, d)$ is unity (absorb all radiation) for particles whose characteristic length $2\pi d$ is larger than wavelength and zero vice-versa as
\begin{eqnarray}
  Q_{\rm PR} (\lambda) &= 0,  \hspace{0.5cm} {\rm if} \hspace{0.1cm}  \lambda > 2\pi d\\
  Q_{\rm PR} (\lambda) &= 1,  \hspace{0.5cm} {\rm if} \hspace{0.1cm}  \lambda \leq 2\pi d
\end{eqnarray}
Here, radiation flux is assumed to be the black body radiation and we use the Planck function $B(\lambda, T_{\rm pla})=C \times (2h c^2/\lambda^5)(1/(e^{hc/\lambda k T_{\rm pla}} - 1))$ whose integral over the $\lambda$ is unity (C is constant). Using the above assumptions, we can express $\bar{Q}_{\rm RP}$ as
\begin{equation}
	\bar{Q}_{\rm RP}(\lambda_{\rm cri},T_{\rm pla} ) = \int_0^{\lambda_{\rm cri}} B( \lambda,T_{\rm pla} ) Q_{\rm RP}(\lambda) d\lambda
\label{Qbar}
\end{equation}
where $\lambda_{\rm cri}=2 \pi d$ \citep{Zoo75}.\\

Figure \ref{beta} shows the $\beta$ value of the above “ideal” particles obtained by equation \ref{betavalue}-\ref{Qbar}. We found that $\beta$ ranges widely from $\sim0.01-50$ depending on density, planet temperature and particle size. Note that, however, if we consider the material properties such as more realistic absorption and scattering properties and calculate $\beta$ more precisely by using the famous Mie theory \citep[e.g.][]{Bur79}, the $\beta$ value may deviate from our ideal results. Thus, in this paper, we use $\beta$ as a parameter ($0.01 < \beta < 10$) and estimate the amount of depletion as will be discussed in the following sections.\\

\subsection{Volatile dust depletion from Martian moon-forming disk} 
\begin{figure}[ht!]
\plotone{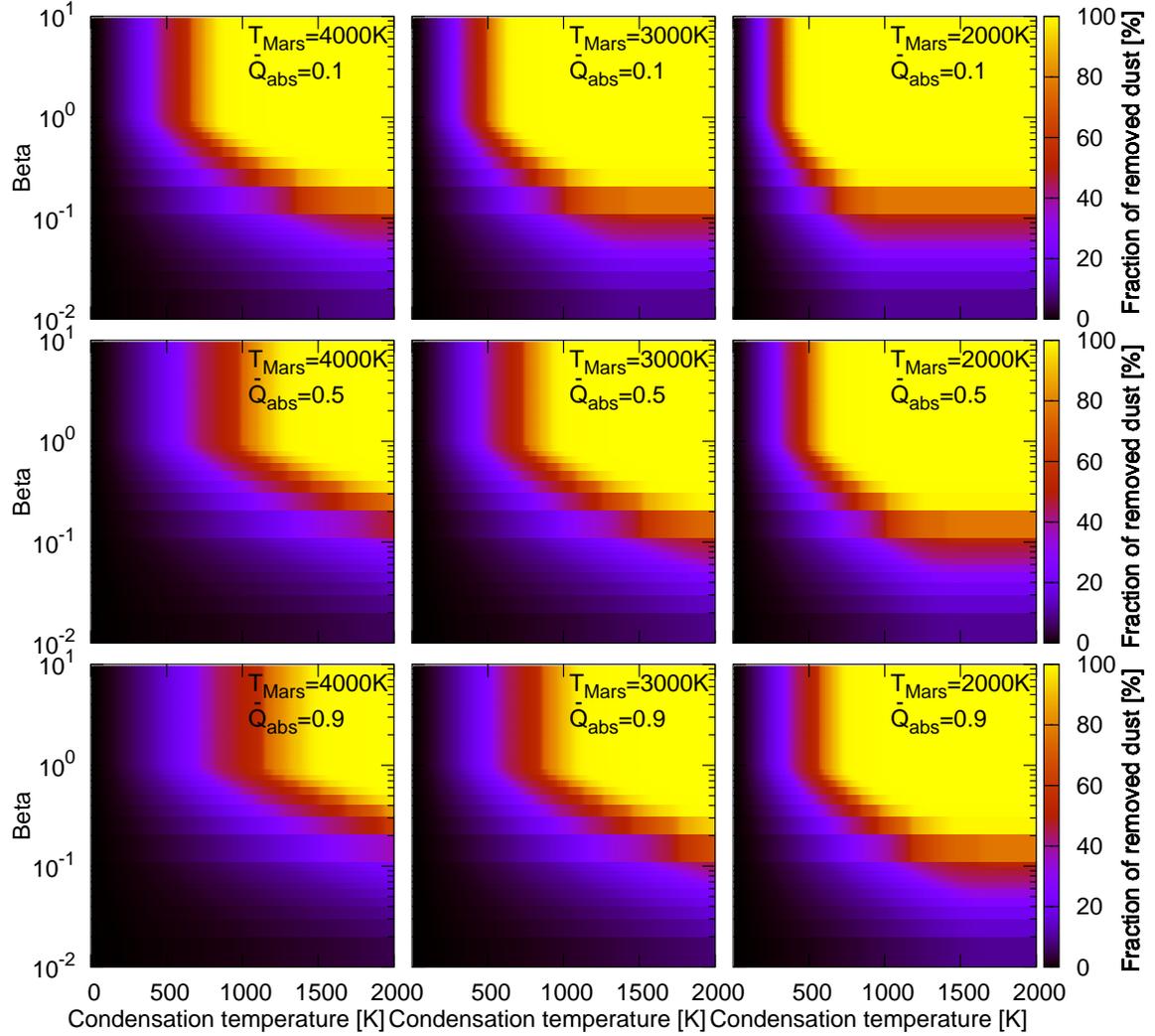}
\caption{Fraction of condensates in Martian moon-forming debris that are removed from Mars' system during their first orbit from their pericenter to apocenter just after the impact as a function of their condensation temperature and $\beta$ value. Different panels show different Mars temperatures and $\bar{Q}_{abs}$. The orbital data is obtained from SPH simulations \citep{Hyo17a,Hyo17b}.}
\label{frac_RP}
\end{figure}

In this subsection, we will estimate the amount of condensed dust depletion from the building blocks of Phobos and Deimos just after the giant impact. As explained before, we will only focus on the first orbit from their pericenter to apocenter. During the successive orbits, the debris quickly forms an optically thick (in the radial direction) and vertically thin equatorial circular disk from which Phobos and Deimos accrete \citep{Hyo17b}. Thus, volatile dust removal would only be efficient during this first orbit.\\

During the first orbit, the temperature of particles can change between $T_{\rm peri} \leq T_{\rm dust} \leq T_{\rm apo}$, where $T_{\rm peri}=2000$ K \citep{Hyo17a} and $T_{\rm apo}=T_{\rm con,apo}$ (see equation \ref{Tapo}), respectively. Using the arguments in Section \ref{sec:RP_cond} and the data obtained from SPH simulations in \cite{Hyo17a}, we calculate the fraction of vapor particles from the building blocks of Phobos and Deimos that (1) condense before reaching their apocenter ($T_{\rm con} < T_{\rm con,apo}$) and that (2) have a larger $\beta$ value than the critical $\beta$ described as
\begin{eqnarray}
  \beta_{\rm cri} =  \left( \frac{T_{\rm pla}}{T_{\rm con}} \right)^2 \left( \frac{R}{4a_{0}} \right)  \bar{Q}^{1/2}_{\rm abs},  \hspace{0.5cm} {\rm if} \hspace{0.1cm}  T_{\rm con} < T_{\rm con,peri}\\
   \beta_{\rm cri} =  \left( \frac{T_{\rm pla}}{T_{\rm con,peri}} \right)^2 \left( \frac{R}{4a_{0}} \right)  \bar{Q}^{1/2}_{\rm abs},  \hspace{0.5cm} {\rm if} \hspace{0.1cm}  T_{\rm con} \geq T_{\rm con,peri}
\end{eqnarray}
where $T_{\rm con,peri}$ is
\begin{equation}
	T_{\rm con,peri} = \sqrt{ \frac{R_{\rm p}}{2r_{\rm per}}} T_{\rm pla}  \bar{Q}^{1/4}_{\rm abs}.
\label{}
\end{equation}
\\

In Section 5.3., we have considered only a specific case of orbital elements. Here, using the data obtained from SPH simulations in \cite{Hyo17a}, we have calculated the fraction of removed dust in Martian moon-forming debris. We assume all dust/particles have the same condensation temperature of $T_{\rm con}$ and a radiation pressure coefficient of $\beta$ with their orbital elements distributed within the range obtained from SPH simulations \citep{Hyo17a,Hyo17b}. Then, we calculate the fraction of particles that satisfy the above two criteria (the fraction of particles whose orbits meet the condition to be removed by radiation pressure with the specific values of $T_{\rm con}$ and $\beta$). Figure \ref{frac_RP} shows the results of the calculations. When the temperature of Mars drops, the vapor cools down more easily closer to their apocenter, and thus dust that has smaller condensation temperature can also be removed. When $\bar{Q}_{abs}$ becomes smaller, the same process occurs as the vapor temperature falls and more volatile elements can condense before reaching their apocenter.\\

Figure \ref{frac_RP} tells us that `moderately' volatile elements such as Na and K ($T_{\rm con} \sim 700-2000$ K) whose $\beta > \sim 0.1$ are more easily removed than highly volatile elements such as H$_{2}$O, Pb and C ($T_{\rm con} < ~ 700$ K). This is because highly volatile elements need to go far enough away from Mars to cool down and condense, but there are few particles in the debris that have such orbital elements. The exact value of $\beta$ strongly depends on condensed species and thus more precise quantitative estimation of the volatile loss by radiation pressure requires detailed study about $\beta$ values. Also, we have to note that the condensates would not be in a pure form, but in a mixture. We will leave this matter to future work.\\

\section{Discussion \& Conclusion} \label{sec:conclusion}
The origin of Martian moons is intensely debated. Recent works have shown that the Martian moons Phobos and Deimos could accrete within an impact generated disk \citep[e.g.][]{Ros16,Hes17,Hyo17a,Hyo17b}. In contrast, (even though it is not shown) it has been suggested that Martian moons could be captured asteroids due to their spectral properties \citep[e.g.][]{Bur78}.\\

Now, JAXA (Japan Aerospace eXploration Agency) is planning the Martian Moons eXplorer (MMX) mission. In this mission a spacecraft will be sent to the Martian moons, perform detailed remote sensing and in-situ analysis, and return samples to Earth. Gamma-ray, neutron-ray, and near-infrared spectrometers will be onboard the MMX spacecraft. The gamma-ray spectrometer can measure major elements such Si and Fe, and the neutron-ray spectrometer can measure H concentration. The near-infrared spectrometer can observe absorption features of hydrated minerals. Thus, major elemental ratios and volatile contents, such as H$_{2}$O, would be critical for understanding the origin of Phobos and Deimos, and also constraining the composition of the impactor that hit Mars, if Phobos and Deimos were formed by a giant impact.\\

Previous works have investigated the expected chemical composition of the building blocks of Phobos and Deimos within the framework of the giant impact hypothesis \citep{Ron16,Hyo17a,Pig18}. Using high-resolution SPH simulations, \cite{Hyo17a} found that the building blocks of Phobos and Deimos contain about an equal mixture of Martian material and impactor material with a temperature of $\sim 2000$ K just after the impact. They also found that the building blocks are about $\sim 5$wt\% vaporized and the rest, about $\sim 95$wt\%, is melted just after the impact. Then, using these results obtained in \cite{Hyo17a}, \cite{Pig18} calculated the condensation sequence of vapor assuming a specific type of impactor composition is equally mixed with Martian materials. They found that the vapor and its condensates contain more volatile elements than melts and its solids. And, its composition significantly differs with varying impactor compositions. However, these works consider the system to be a closed one and did not take into account any possible processes that may cause volatile elements to be lost from the Martian system.\\

In this work, we consider hydrodynamic escape (Section 4) and radiation pressure (Section 5) as possible mechanisms to remove volatiles, since the Borealis basin-forming impact would have heated the Martian surface up to $\sim1000-6000$ K (Section 2) and the building blocks of Phobos and Deimos should be heated up to $\sim 2000$ K \citep{Hyo17a}. We focus on the epoch just after the impact because the orbits of the building blocks of Phobos and Deimos are highly eccentric at this time \citep{Hyo17b}, and thus are expected to escape more easily from the system closer to their apocenters where Martian gravitational attraction becomes weaker.\\

In section 4, we consider volatile vapor loss by hydrodynamic escape. Mars has weaker gravity than the Earth and thus the escaping parameter $\lambda_{\rm esc}$ is smaller. If $\lambda_{\rm esc} < 1$, the vapor thermal velocity exceeds the Martian escape velocity and vapor loss would occur. Using the orbital data obtained in  \cite{Hyo17a}, we calculated the fraction of vapor that satisfies $\lambda_{\rm esc} < 1$ during its first orbit from pericenter (which is around the impact point) to apocenter. We found that about $\sim10-40$\% of the vapor would be lost ($\lambda_{\rm esc} < 1$) at $T_{\rm gas} \sim 1000-2000$ K depending on the impactor composition (the mean molecular mass of the vapor depends on the impactor composition).\\

Meanwhile, during the first orbit from pericenter to apocenter, some of the vapor may condense and form $\sim 0.1 \micron$ sized dust \citep{Hyo17a}. And these small dust grains may be influenced by radiation from Mars when it is heated up to $1000 - 6000$ K just after the impact (Section 2). In section 5, we calculated the fraction of $\sim 0.1 \micron$ sized dust that can potentially be blown off by radiation pressure as a function of different $\beta$ values and condensation temperatures during its first orbit. Both $\beta$ and condensation temperature strongly depend on dust composition. We found that removal of `moderately' volatile dust ($700 {\rm K} < T_{\rm con} < 2000$ K) by radiation pressure was more likely to occur when it satisfies $\beta > \sim 0.1$ than “highly” volatile dust ($T_{\rm con} < 700$ K). Further investigation will be required to study condensation temperature and $\beta$ values for different elements and to constrain the exact amount of depletion of volatiles using a chemistry and radiative transfer code.\\

In this work, we qualitatively demonstrated that hydrodynamic escape and radiation pressure can remove volatiles from the building blocks of Phobos and Deimos just after the impact. Therefore, not only the bulk chemical composition, but also the bulk volatile elements content would be key measurements to distinguish the two hypotheses for the origin of Phobos and Deimos - capture scenario or impact scenario. Thus, the absence of volatiles obtained in JAXA's MMX or variation in their content from the one predicted in our previous works \citep[][where a closed system was considered]{Hyo17a,Pig18} could further confirm the impact origin of the Martian moons and also tell us the efficiency of the processes considered here.

\acknowledgments
We thank Vincent Bourrier for discussion on radiation pressure. R.H acknowledge the financial supports of JSPS Grants-in-Aid for JSPS Fellows (JP17J01269). S.C., R.H., and H.G. acknowledge the financial support of the JSPS-MAEDI bilateral joint research project (SAKURA program). R.H. and H.G. thank the Astrobiolgy Center of the National Institutes of Natural Sciences, NINS (AB291011). H.G. also acknowledges JSPS KAKENHI grant (JP17H02990) and MEXT KAKENHI grant (JP17H06457).



\end{document}